\documentclass[aps,reprint,floatfix,superscriptaddress,showpacs,twocolumn,nofootinbib,longbibliography]{revtex4-1}

\usepackage{graphicx}
\usepackage{amsmath}
\usepackage{amssymb}
\usepackage{amsfonts}
\usepackage{amsthm}
\usepackage{mathtools}
\usepackage{siunitx}
\usepackage[unicode=true, breaklinks=true, pdfborder={0 0 1}, backref=false, colorlinks=true, linkcolor=blue, citecolor=blue, urlcolor=blue]{hyperref}
\usepackage{color}
\usepackage{soul}
\usepackage{textcomp}
\usepackage{bm}
\usepackage{dsfont}
\usepackage{relsize}
\usepackage{braket}
\usepackage{enumitem}   
\usepackage{mathrsfs}
\usepackage{cleveref}
\usepackage{bigints}
\usepackage{framed}
\usepackage[makeroom]{cancel}
\usepackage{comment}
\usepackage{notes2bib}

\theoremstyle{plain}

\numberwithin{obs}{section}

\definecolor{Blue}{rgb}{0,0,0}
\definecolor{Red}{rgb}{1,0,0}
\definecolor{Green}{rgb}{0,1,0}
\definecolor{darkgreen}{rgb}{0,.7,0}
\definecolor{Orange}{rgb}{1,0.5,0}
\definecolor{Purp}{rgb}{.2,0,.2}
\definecolor{white}{rgb}{1,1,1}

\newcommand{\marti}[1]{\textcolor{Blue}{#1}}

\definecolor{ms}{rgb}{0,.4,1}

\newcommand{\bigo}[1]{\mathcal{O}\left (#1\right)}

\newcommand{\average}[2]{\left \langle #1 \right \rangle_{#2}}
\newcommand{\rom}[1]{\uppercase\expandafter{\romannumeral #1\relax}}
\newcommand{\norbra}[1]{\left( #1\right)}

\usepackage{amssymb}             
\newcommand{\dt}{{\rm d}t\,}
\newcommand{\de}{{\rm d}}

\newcommand{\Tr}[1]{\mathrm{Tr}\left[ #1\right]} 

\begin{document}
	\title{ 
	\marti{Minimally dissipative information erasure in a quantum dot via thermodynamic length}
	}
	\author{Matteo Scandi}
	\affiliation{ICFO - Institut de Ciencies Fotoniques, The Barcelona Institute of Science and Technology, Castelldefels (Barcelona), 08860, Spain}
	\author{David Barker}
	\affiliation{NanoLund and Solid State Physics, Lund University, Box 118, 22100 Lund, Sweden}
	\author{Sebastian Lehmann}
	\affiliation{NanoLund and Solid State Physics, Lund University, Box 118, 22100 Lund, Sweden} 
	\author{Kimberly~A.~Dick}
	\affiliation{NanoLund and Solid State Physics, Lund University, Box 118, 22100 Lund, Sweden}
	\affiliation{Centre for Analysis and Synthesis, Lund University, Box 124, 22100 Lund, Sweden.}
	\author{Ville F. Maisi}
	\affiliation{NanoLund and Solid State Physics, Lund University, Box 118, 22100 Lund, Sweden}
	\author{Mart\'{i} Perarnau-Llobet}
	\affiliation{D\'{e}partement de Physique Appliqu\'{e}e, Universit\'{e} de Gen\'{e}ve, Gen\'{e}ve, Switzerland}

	
	\begin{abstract}
	    In this work we explore the use of thermodynamic length to improve the performance of experimental protocols. In particular, we implement Landauer erasure on a driven electron level in a semiconductor quantum dot, and compare the standard protocol in which the energy is increased linearly in time with the one coming from geometric optimisation. The latter is obtained by choosing a suitable  metric structure, whose  geodesics correspond to optimal finite-time thermodynamic protocols in the slow driving regime.  We show experimentally that geodesic drivings minimise dissipation for slow protocols, with a bigger  improvement as one approaches perfect erasure.  Moreover,  the geometric approach also leads to smaller dissipation even when the time of the protocol becomes comparable with the equilibration timescale of the system, i.e., away from the slow driving regime. 
		Our results also illustrate, in a single-electron device, a fundamental principle of thermodynamic geometry: optimal finite-time thermodynamic protocols are those with constant dissipation rate along the process. 
	\end{abstract}

	\maketitle
	
	
	Landauer erasure represents one of the most paradigmatic protocols in stochastic and quantum thermodynamics. 
Its relevance is not only historical, as it was the first case in which a strong argument for the physicality of information was made, but also conceptual, as it shows how logical irreversibility inevitably leads to dissipation, and practical, as it imposes a fundamental bound on the minimal heat released by an operating computer with finite memory. In particular, Landauer's limit establishes that the minimal amount of heat dissipated in order to erase a bit is bounded by~\cite{landauer1961irreversibility}:
	\begin{align}
		\Delta Q \geq -k_B T \,\Delta S \label{eq:LandauerLimit}
	\end{align}
	where $T$ is the temperature of the bath and $\Delta S$ is the difference in entropy between the final and the initial state, which turns out to be negative for erasing protocols. 

	
	Equality in Eq.~\eqref{eq:LandauerLimit} corresponds to an ideal isothermal process. This can only be realised in infinite time, which makes Landauer's limit \emph{de facto} unattainable in practice. Nevertheless, it is a crucial task to minimise  dissipation (i.e. $\Delta Q$) in information-processing devices, and much experimental effort has been devoted to approach the Landauer's limit~\cite{Ciliberto2017}. Experimental demonstrations of (almost-perfect) Landauer erasure have been reported in different  platforms, including colloidal particles~\cite{berut2012experimental,jun2014high,Brut2015,gavrilov_erasure_2016}, 
	nanomagnets~\cite{Hong2016,Martini2016,Gaudenzi2018}, superconducting flux logic cells~\cite{saira_nonequilibrium_2020},  underdamped micromechanical
	oscillators~\cite{Dago2021,dago_dynamics_2022} and optomechanical systems~\cite{ciampini2021experimental} (see also related works in quantum systems such as  nuclear magnetic resonance set-ups~\cite{peterson2016experimental} and ion traps~\cite{Yan2018}). 
	
	
	Despite how well studied this problem is, in all the experimental works above  the driving chosen in order to induce the erasure is linear in time. We show here that this is suboptimal, which is in agreement with previous theoretical works~\cite{Aurell2011a,Diana2013,Zulkowski2014,scandiThermodynamicLengthOpen2019,Proesmans2020,Proesmans2020II,Zhen2021,VanVu2022,Zhen2022,Yu-Han2022,lee2022speed}.  In particular, we study how to exploit the concept of thermodynamic length~\cite{Salamon3,Salamon1,Nulton1985, Andresen,Crooks,Zulkowski,Sivak2012a,Bonana2014,scandiThermodynamicLengthOpen2019,abiusoGeometricOptimisationQuantum2020,Deffner2020a} to devise better erasing protocols in finite time. This quantity arises from the expansion of the entropy production for protocols that are performed in a long, but finite time. In this regime, optimal protocols are governed by the principle of constant dissipation rate,
	meaning that the optimal protocol is the one that allocates the dissipation in the most uniform way~\cite{salamon_minimum_1980,salamon_principles_2001,andresen_current_2011,abiusoGeometricOptimisationQuantum2020}. The corresponding thermodynamic metric also gives a prescription to find minimally dissipating drivings: in fact, the geodesics associated to this metric realise optimal protocols in the slow driving regime~\cite{Salamon3, Salamon1,Nulton1985,Sivak2012a,scandiThermodynamicLengthOpen2019}.
	
\begin{figure}
		\centering
		\includegraphics[width=\columnwidth]{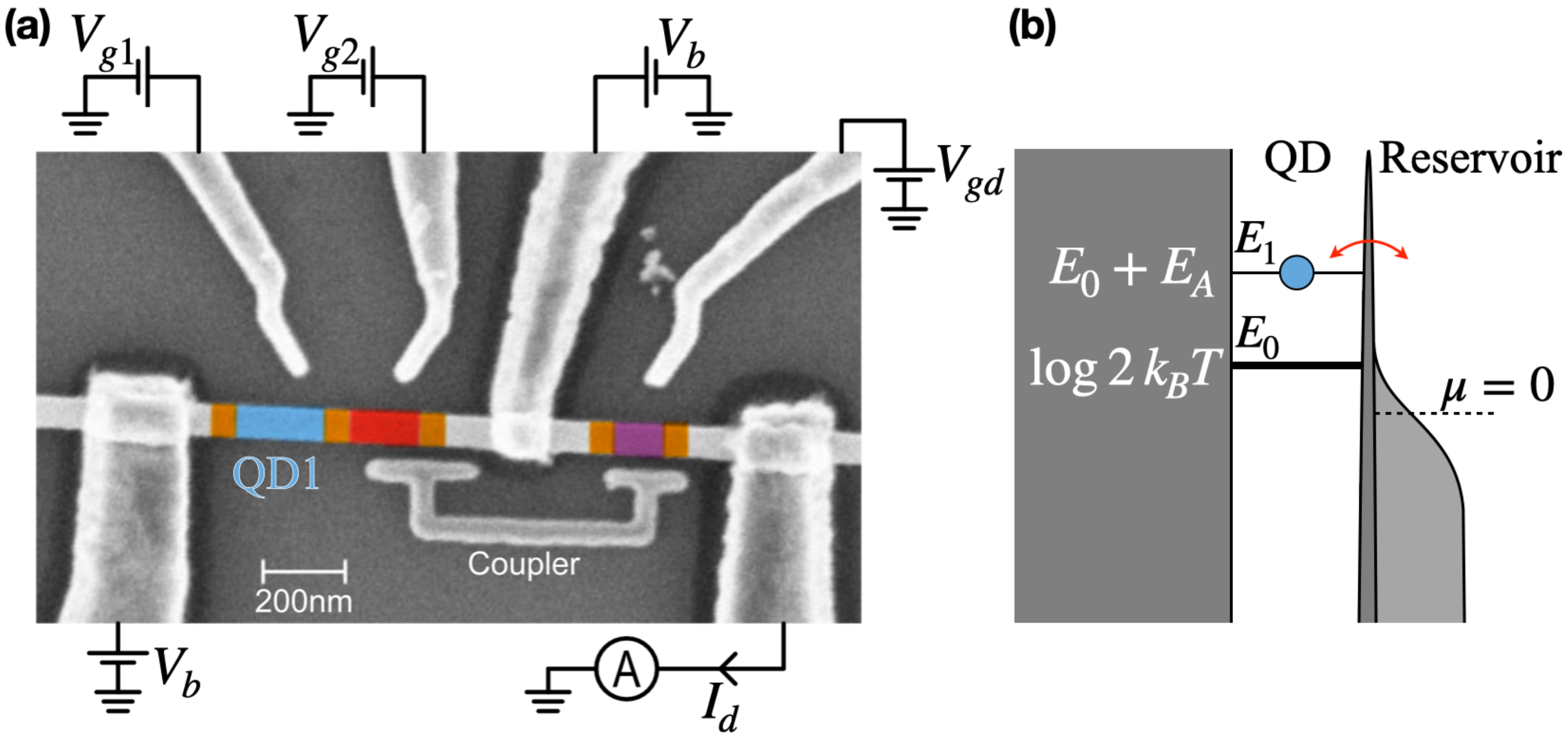}
		\vspace{-2em} 
		\caption{a) Scanning electron microscope image of the nanowire device. Embedded in the nanowire are three QDs, each aligned to one of the plunger gates $V_{g1}$,$V_{g2}$ or $V_{gd}$. Contacts separate the device into one part with two QDs and one part with a single QD. The coupler couples the two systems together, allowing the current $I_d$ through the lone QD to provide a measure of the charge state of the other system. Here, the QD involved in the experiment is marked in blue (close to the plunger gate with $V_{g1}$) while the quantum dot marked in red is tuned into Coulomb blockade. The sensor quantum dot is marked in purple and the tunnel barriers are coloured orange. b) The energy diagram for the protocol.}
		\label{fig:device}
	\end{figure}

	We experimentally demonstrate how this geometric approach can be exploited to minimise  dissipation in a Landauer erasure protocol. Our device is based on a semiconductor quantum dot which allows for the manipulation of discrete energy levels, see Fig.~\ref{fig:device}. We study both the regime of slow driving, for which we demonstrate the expected improvement, and the fast driving regime. For the latter, 
	which is in principle outside of the realms of application of thermodynamic length, we still observe substantial reductions in dissipation compared to the linear drive. 
	Finally, we show that the improvements become more and more relevant the closer one gets to complete erasure.
	
	These results can be regarded as the experimental proof of principle for the relevance of thermodynamic length in devising optimal finite-time protocols. As it was argued theoretically  in~\cite{salamon_minimum_1980,salamon_principles_2001,andresen_current_2011,Sivak2012a,scandiThermodynamicLengthOpen2019,abiusoGeometricOptimisationQuantum2020,Deffner2020a},  thermodynamic length offers a flexible and powerful tool for minimising dissipation.  It is particularly interesting to see that even for a problem as well explored as the one of Landauer erasure it is possible to find an improvement over  present experimental protocols.
	

\emph{Experimental set-up}. 	The experiment is performed using the same device as in~\cite{barker2022experimental}, shown in Figure~\ref{fig:device} (a). Three quantum dots (QDs) are formed by polytype engineering in an InAs nanowire~\cite{lehmann_general_2013,nilsson_transport_2016,chen_conduction_2017,barker_individually_2019}. The occupancy $n \in \{0,1\}$ of a spin-degenerate energy level $E$ in the leftmost QD (QD1) encodes the bit of information to be erased in the experiment. We drive the energy level with the plunger gate voltage $V_{g1}$ which has a lever arm $\alpha = 1.6\times10^4\,k_BT/\mathrm{V}$. The rightmost QD is voltage biased with $V_b = \SI{0.5}{\milli\volt}$ and tuned so that the current $I_d$ is sensitive to changes in the QD1 occupancy, giving a real-time probe of $n$~\cite{vandersypen_realtime_2004,gustavsson_electron_2009}. The middle QD is kept in Coulomb blockade, reducing the system to the one shown in Figure~\ref{fig:device} (b): a discrete QD1 energy level coupled to a fermionic reservoir at temperature $T = \SI{100}{\milli\kelvin}$ (set by the cryostat temperature). Electrons tunnel between them with the rates $\Gamma_{\mathrm{in}}=2\Gamma_0(1+bE)f(E)$ and $\Gamma_{\mathrm{out}}=\Gamma_0(1+bE)(1-f(E))$ where $\Gamma_0 = \SI{39}{\hertz}$ and $b=0.0036/k_BT$ were determined using a feedback protocol~\cite{hofmann_measuring_2016}, and $f(E)=1/(1+e^{E/k_BT})$ is the Fermi-Dirac distribution. The average occupation at equilibrium for each energy is given by 
	$n_{eq}(E)  := 1/(1 + \frac{1}{2}\, e^{E/k_BT})$,
	corresponding to the thermal state for a system with a degeneracy 2 in the $n=1$ state.

	\emph{Erasure protocol} The erasure protocol is realised as follows: first, the system is allowed to thermalise in contact with the reservoir bath while keeping its energy at $E_0 = \log 2 \,k_B T$ corresponding to a $50\, \% - 50\, \%$ occupation condition, see Fig.~\ref{fig:device}~(b). Then, while  still keeping it in contact with the bath, we ramp up the energy of the dot until we reach $E_1 := E_0 + E_A$, where $E_A$ defines the driving amplitude. When $E_A\gg k_B T$, we have $n_{eq}(E_1)\approx 0$, i.e., the dot is  unoccupied  with probability close to 1. As the last step, the energy is quenched back to $E_0$, so that the system is effectively erased. 
	
	We measure the heat $\Delta Q$ by monitoring the electron transitions: whenever the dot is occupied and an electron tunnels out, the energy at that time gets transferred to the reservoir where it dissipates and adds to $\Delta Q$; similarly, if an electron tunnels into the dot that energy is subtracted. Moreover, since the quench is instantaneous, it does not contribute to the heat production, as the state of the system is unaffected by it. Repeating the protocol many times one can then compute the average heat $\average{\Delta Q}{}$ simply by adding up the resulting heat for each round and dividing by the number of rounds. Alternatively, the same result can also be obtained from the average occupation $\average{n(t)}{}$ thanks to the equality:
	\begin{align}
		\average{\Delta Q}{} = - \int_0^\tau \dt\, \average{\dot n(t)}{} E(t),\label{eq:heatPopulation}
	\end{align}
	where $\tau$ is the total time of the protocol, while $E(t)$ is the drive used to interpolate between $E_0$ and $E_1$. The most usual choice is to take it to be a linear drive $E(t) := E_0 + E_A\cdot\,t/\tau$, but in principle $E(t)$ could be any function satisfying $E(0) = E_0$ and $E(\tau)=E_1$.  In fact, it turns out that the linear protocol is suboptimal.
	
	\begin{figure}
		\centering
		\includegraphics[width=\columnwidth]{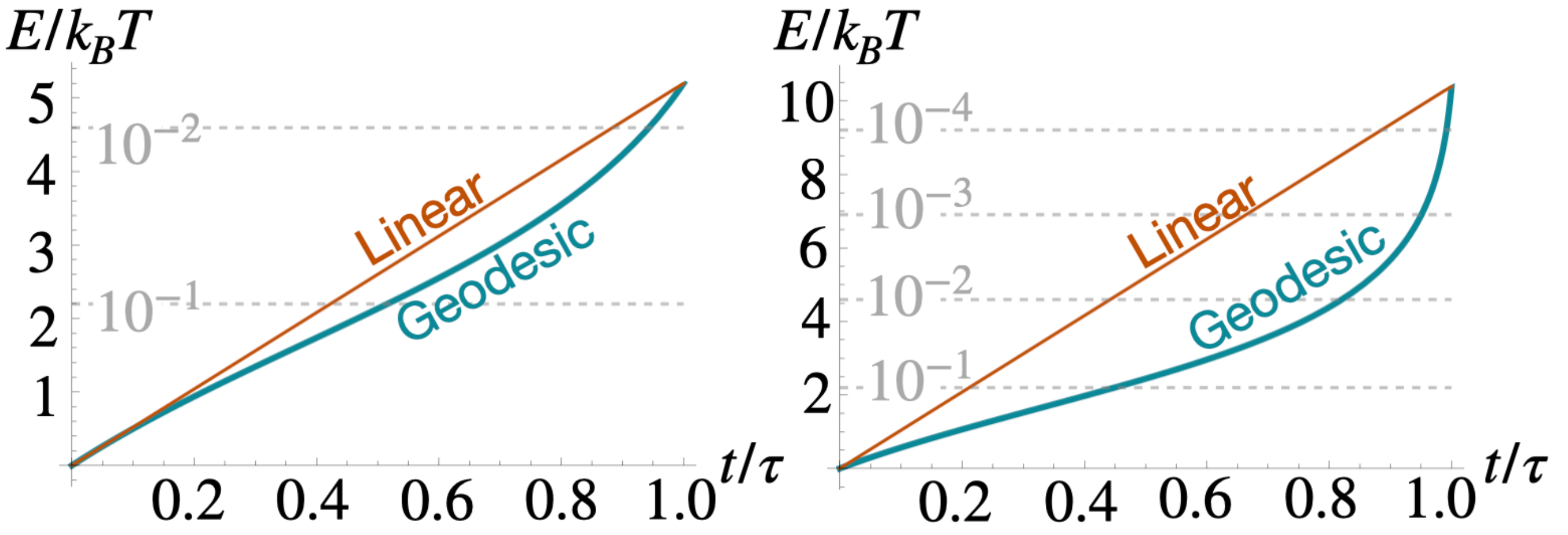}
		\vspace{-1em} 
		\caption{Comparison between the linear drive and the geodesic drive for amplitudes $E_A = 5.2 \,k_BT$ and $E_A = 10.4\, k_BT$. The dotted lines in grey indicate the equilibrium population of the excited state for each energy. As we can see, the geodesic drive allocates more time for the ramping up when the excited state is more occupied, while it becomes steeper at the end of the protocol. }
		\label{fig:trajectories}
	\end{figure}
	
	An alternative protocol can be designed as follows. First, it should be noticed that in the limit of $(\Gamma_0 \tau)\gg 1$, Eq.~\eqref{eq:heatPopulation} can be brought to the form~\cite{supmat}:
	\begin{align}
		\average{\Delta Q}{} &=-k_BT \,\Delta S + k_BT \int_0^\tau \dt\, g(t) \dot E(t)^2\label{eq:metricForm}
	\end{align}
	up to corrections of order $(\Gamma_0 \tau)^{-2}$ and regardless of the particular choice of the protocol. The quantity $g(t)$ is called thermodynamic metric: it is always positive and depends smoothly on the drive $E(t)$. For reasons of space, we refer for the particular expression of the metric to the Supplemental Material (SM)~\cite{supmat}. The integral in Eq.~\eqref{eq:metricForm}  is a standard quantity in differential geometry, usually called energy functional. The name comes from the analogy with the action of a particle moving with velocity $\dot E(t)$ and variable mass $g(t)$. Interestingly, thanks to the form of the dissipation in Eq.~\eqref{eq:metricForm}, we can automatically construct minimally dissipating drives simply by solving the geodesic equation for $E(t)$. Further details are provided in the SM~\cite{supmat} and in~\cite{scandiThermodynamicLengthOpen2019,abiusoGeometricOptimisationQuantum2020}.
	
	The corresponding trajectory are shown in Fig.~\ref{fig:trajectories} for two driving amplitudes. Compared with the linear drive, the geodesic one allocates more time to ramp up the energy when the QD is occupied with larger chance (at low $E$) and becomes steeper towards the end of the protocol. This can be intuitively understood as follows: since the dissipation is linear in $\average{ \dot n(t)}{}$ while $\average{n(t)}{}$ decreases exponentially with $E(t)$, it is better to allocate more time at the beginning, when the variation $\average{ \dot n(t)}{}$ is big, and to reserve little time to the final jump in the energy, because exponentially small amount of the tunneling events take place at large $E$. Notice that this reasoning is justified by the fact that for slow driving $\average{  n(t)}{}\simeq n_{eq}(E(t))$. Still, we show that this intuition is also relevant for drives where $\Gamma_0\tau \simeq 3 $ (which we dub fast driving regime).
	
\begin{figure}
		\centering
		\includegraphics[width=\columnwidth]{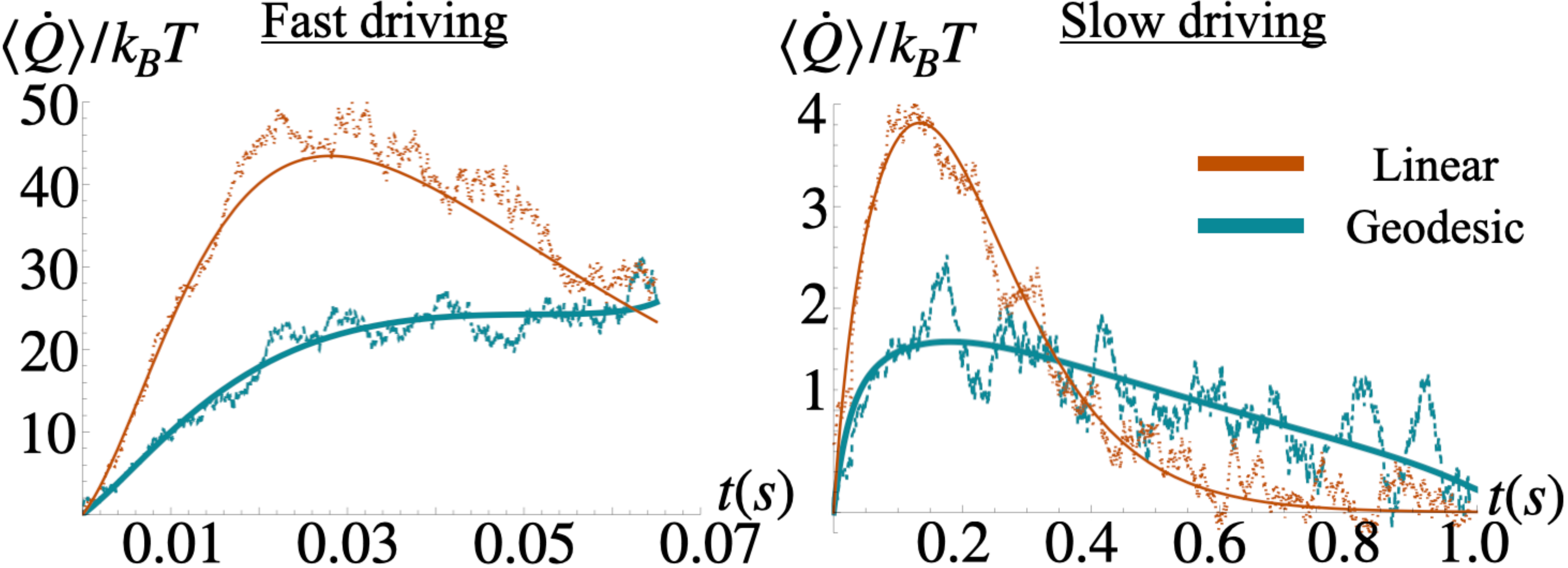}
		\vspace{-1em} 
		\caption{Entropy production during Landauer erasure for amplitude $E_A = 10.4\, k_BT$ and erasure times of $\tau = 0.07\, s$ and $\tau = 1.0 \,s$. The continuous lines are the theoretical predictions, while the dotted lines are the experimental data. The derivative has been performed by regularising the experimental data through a mean filter. }
		\label{fig:entropyproduction}
	\end{figure}

	The reasoning above intuitively captures a characteristic of geodesic drives: it can be proven that the entropy production rate, i.e., the integrand in Eq.~\eqref{eq:metricForm}, is constant along optimal protocols~\cite{salamon_minimum_1980,salamon_principles_2001,andresen_current_2011,abiusoGeometricOptimisationQuantum2020}. This effect is exemplified in Fig.~\ref{fig:entropyproduction}, where we plot the heat  production rate both in the fast and in the slow driving regime ($\Gamma_0\tau \simeq 40 $ for the latter), comparing the behaviour of a linear drive with the one of the geodesic. We see that for the non-optimized drive, the heat production peaks at the beginning, while decreasing towards zero at the end of the protocol. For geodesic drives instead, the heat is produced more uniformly along the protocol\footnote{The fact that the entropy production rate is not perfectly constant along the trajectory arises from finite time effects. We numerically verified that increasing $\Gamma_0\tau $ makes the heat production closer to a constant value.}.

\begin{figure}
		\centering
		\includegraphics[width=\columnwidth]{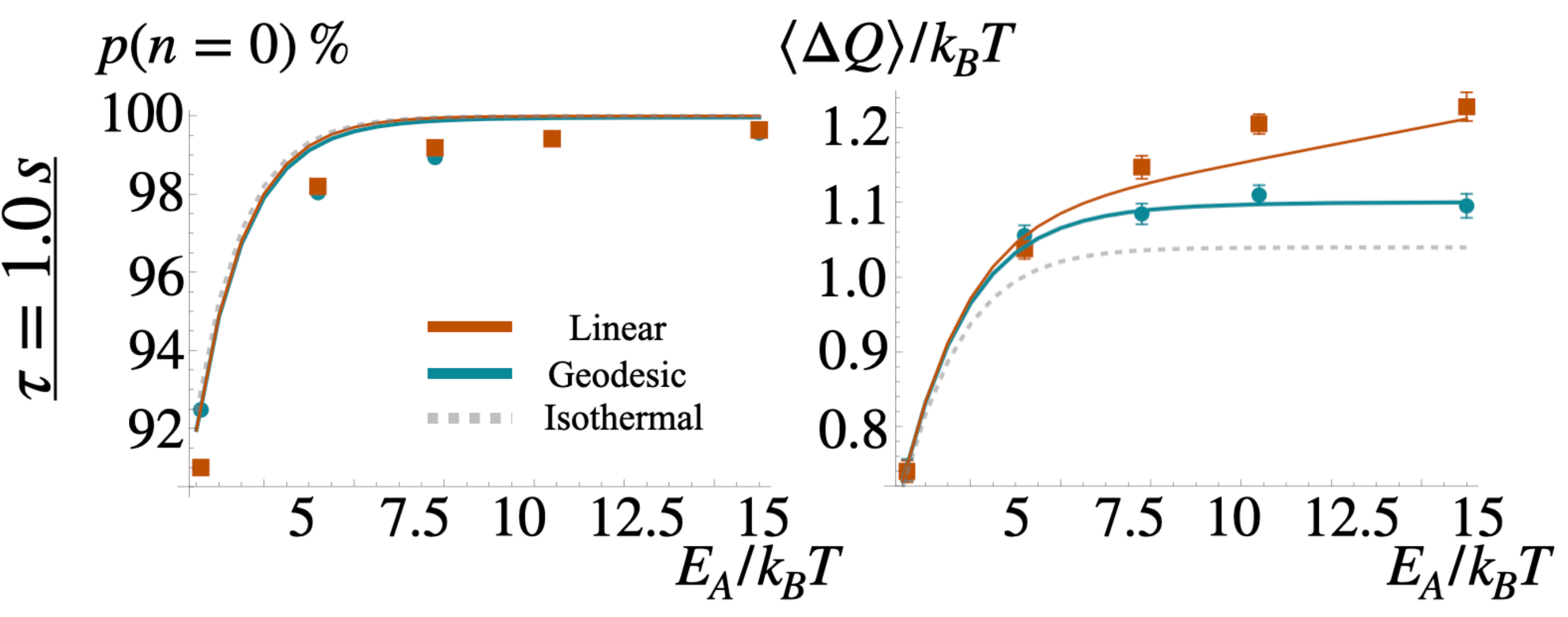}
		\includegraphics[width=\columnwidth]{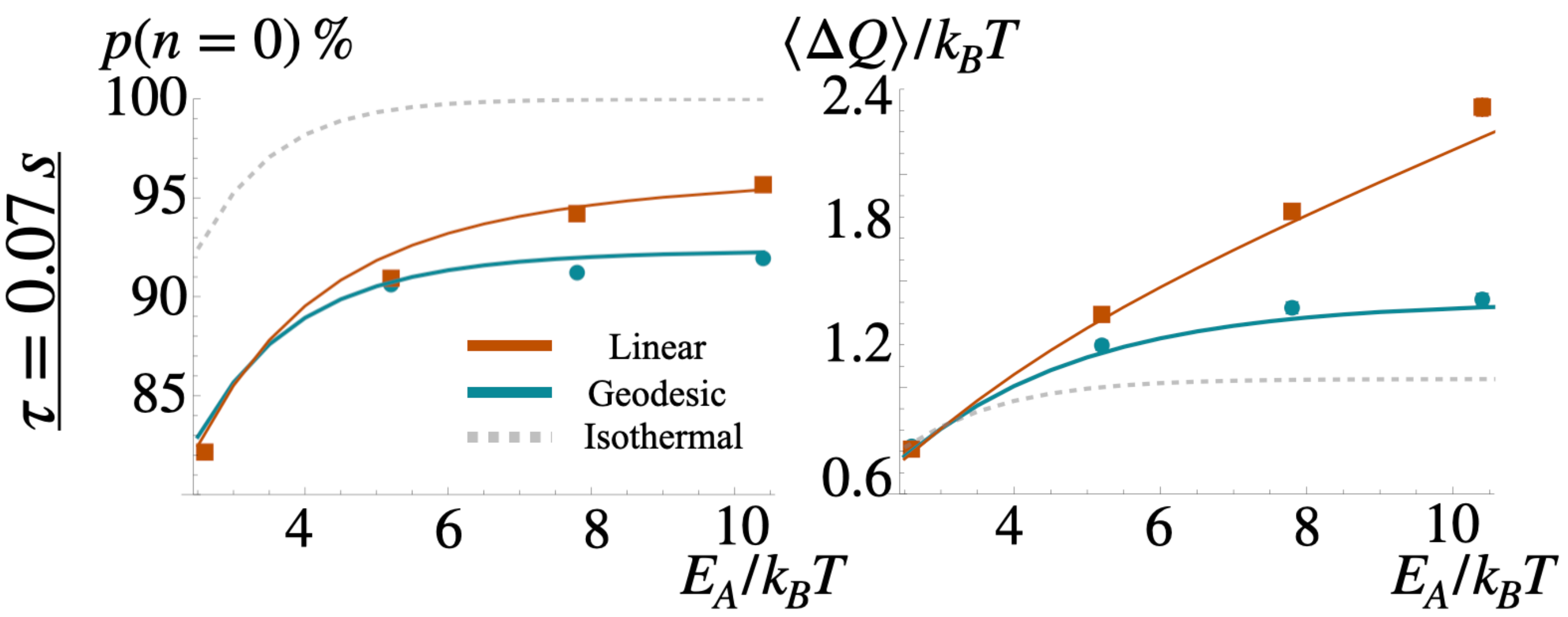}
		\vspace{-2em} 
		\caption{In the two left panels we plot the heat as a function of the erasure quality of the protocol (given by $1-\average{n(\tau)}{} = p(n=0)$, the average population in the empty state). The panel above refers to $\tau =1.0\,s$,  while the bottom one corresponds to $ \tau =0.07 \,s$ (same convention used for the right panels). On the right, we plot the heat produced as a function of the driving amplitude. The continuous lines are the theoretical predictions, while the dotted lines corresponds to the experimental data. The grey dotted line corresponds to the ideal case, that is the maximal erasure for each $E_A$ on the left, and the Landauer's limit $k_BT \Delta S$ on the right.}
		\label{fig:data1}
	\end{figure}

\emph{Comparison between linear and geodesic drive}. 	In this section we compare the performance of the geodesic protocol with the usual choice of a linear drive. The data are presented in Fig.~\ref{fig:data1}.
	
	The two plots on the left represent the quality of erasure as a function of the driving amplitude. This quantity is measured by the percentage of residual population in the dot at the end of the drive or, equivalently, with the population probability $p(n=0)$. On the right of   Fig.~\ref{fig:data1}, we also plot the dissipated heat as a function of the driving amplitude $E_A$. The above plots refer to the 
	slow	driving ($\tau = 1.0\,s$, $(\Gamma_0\tau) = 39$), while the two on the bottom refer to the fast driving regime ($\tau = 0.07\,s$, $(\Gamma_0\tau)= 2.73$). \marti{ These experimental results are complemented by numerical simulations in the SM~\cite{supmat}, where we confirm that the point $\tau = 1.0\,s$ is deep in the slow driving regime, whereas for $\tau = 0.07\,s$ the slow driving approximation~\eqref{eq:metricForm} breaks down. }
	
	When the amplitude is small, we see that the results given by the geodesic drive are similar to the ones for the linear drive (we even find a marginally better erasure in the $1\,s$ drive). This can be explained by the fact that for small amplitudes the geodesic does not depart much from the linear drive, see Fig.~\ref{fig:trajectories}. For larger amplitudes, we  can appreciate the strength of geodesic protocols.  In the slow driving regime (top figures), we observe that the geodesic drive dissipates less (right top Fig.~\ref{fig:data1}) for a similar quality of erasure (left top Fig.~\ref{fig:data1}). In fact, the dissipation grows linearly as the amplitude $E_A$ increases for the linear protocol (i.e., as the quality of the erasure increases), whereas it tends to a constant for the geodesic drive. This makes geodesic drives  more and more relevant when one wants to reach higher erasing quality. Indeed, the geodesic drive stays much closer to the Landauer's limit of $k_BT\Delta S$. These results demonstrate the reduction in dissipation when erasing a qubit in  the slow driving regime theoretically predicted in previous works~\cite{scandiThermodynamicLengthOpen2019}. \marti{ In the SM~\cite{supmat}, we complement these experimental results with numerical simulations that show the reduction of dissipation achieved through the geodesic drive as a function of $\tau$.    }

	Interestingly, as we depart from the slow driving regime (bottom Fig.~\ref{fig:data1}), we observe	a trade-off: on the one hand, the fast linear drive achieves a higher quality of erasure than the geodesic protocol. This happens because for such a short protocol duration, the system does not have enough time to respond to the steep ramp at the end of the geodesic drive making that energy range effectively lost in the erasure attempt. On the other hand, the dissipation produced by the geodesic protocol saturates, as expected from the theory~\cite{supmat}, so one can achieve the same erasing precision as the one given by the linear drive at the same dissipation just by increasing the amplitude. 
	
	It should be noticed that if one allows for a small extra time at the end of the protocol in which the system thermalizes at a fixed energy, the difference in the quality of erasure between the linear and the geodesic drive would disappear. On the other hand, since the biggest contribution to the dissipation comes from the initial part of the protocol, if one can allow for this additional time, this would make the geodesic drive preferable because it would give the same erasure quality at lower dissipation. This intuition is made precise in the SM~\cite{supmat} via numerical simulations of the process. In this way, there is a trade-off between the precision of erasure and time at optimal dissipation, or between dissipation and quality of erasure for a fixed time.  
	
	
	
\emph{Conclusions. } 	The present work shows the relevance of thermodynamic length in the design of minimally dissipating experimental protocols. In particular, by considering the erasure of information in a quantum dot we showed that even in such a well studied protocol a simple application of our method decreases the amount of dissipation released during the driving. 
	This
	comes at the cost of a small decrease in the quality of the erasure, which can arguably be recovered by allowing a small transient at the end of the transformation, or by moving to bigger driving amplitudes (thanks to the saturation of the dissipated heat shown on the right of Fig~\ref{fig:data1}). Moreover, we showed that even if the thermodynamic length in principle should only apply to the slow driving limit, it improves also on relatively fast protocols, proving the wide applicability of this approach.
	
	This universality comes from an underlying physical principle: the dissipation rate in optimal protocols should be constant along the trajectory~\cite{salamon_minimum_1980,salamon_principles_2001,andresen_current_2011,abiusoGeometricOptimisationQuantum2020}. In this sense, what the geodesic drive does is allocating the heat production in a more uniform way compared with the one arising from the naive choice of a linear drive. This fact is of key importance in the interpretation of the shape of the geodesic protocols and provides an intuitive method to develop optimal drives.
	
	Beyond the minimisation of average dissipation, the geodesic drives considered here can also become useful for the minimisation of work and heat fluctuations~\cite{Crooks,Miller2019}, for probabilistic work extraction~\cite{Maillet2019optimal,miller2022finite}, and for increasing the efficiency of thermal machines~\cite{abiuso2020optimal,Brandner2020,Miller2020geom,frim2021geometric}. Future works include the implementation of optimal protocols in the fast driving regime~\cite{Proesmans2020,Proesmans2020II,Blaber2021,Zhen2021,Zhen2022} and observing effects arising from  quantum coherence in erasure processes~\cite{Scandi2020,miller2020quantum,VanVu2022}.

	~\\
	\emph{Acknowledgments.}  M. S. acknowledges support from the European Union’s Horizon 2020 research and innovation programme under the Marie Sk\l{}odowska-Curie grant agreement No 713729, and from the Government of Spain (FIS2020-TRANQI and Severo Ochoa CEX2019-000910- S), Fundacio Cellex, Fundaci\'{o} Mir-Puig, Generalitat de Catalunya (SGR 1381 and CERCA Programme). D. B. and V. F. M. thank for financial support from NanoLund, Swedish Research Council (Dnr 2019-04111), grant number FQXi-IAF19-07 from the Foundational Questions Institute, a donor advised fund of Silicon Valley Community Foundation, and the Knut and Alice Wallenberg Foundation (KAW) via Project No. 2016.0089. M. P.-L.  acknowledges funding from Swiss National Science Foundation through an Ambizione grant PZ00P2-186067. 
	
	\bibliography{Bib}
	\onecolumngrid
	\appendix
	
	\section{Experimental set-up}
	
	
	\subsection{Tunnel rates and lever arm}
	\noindent \marti{The device used in the erasure experiment is the same as was used in Ref.~\citealp{barker2022experimental}, and the data processing was performed in the same manner as described in the Supplemental Material of that reference. However, the present work uses a different charge transition which means the tunnel rates needed to be measured again for this specific study. The measurement was performed using a feedback method developed in Ref.~\citealp{hofmann_measuring_2016}, and the results are shown in Supplemental Fig.~\ref{fig:tunnelrates}~(a). From this measurement, the lever arm $\alpha$ was extracted by using the detailed balance relation $\Gamma_{\mathrm{in}}/\Gamma_{\mathrm{out}} = 2\exp(\alpha V_{g1}/k_BT)$, allowing for the conversion between a change in the gate voltage $V_{g1}$ and energy $E$ as $E = -\alpha \Delta V_{g1}$. Supplemental Figure~\ref{fig:tunnelrates}~(b) presents the measured data showing that the detailed balance holds for $|\Delta V_{g1}|\leq 0.35$ mV. Fitting a straight line to gate voltage range yields $\alpha = 1.6\times10^4\; k_BT/\mathrm{V}$.
		As a second step, the measured $\Gamma_{\mathrm{in}}$ and $\Gamma_{\mathrm{out}}$ of Supplemental Fig.~\ref{fig:tunnelrates}~(a) were simultaneously fitted to $\Gamma_{\mathrm{in}} = 2\Gamma_0(1+b E)f(E)$ and $\Gamma_{\mathrm{out}} = \Gamma_0(1+b E)(1-f(E))$, where $f(E)$ is the Fermi-Dirac distribution in the electronic reservoir. The fits yield the fitting parameter values $b = 0.0036/k_BT$ and $\Gamma_0 = 39$~Hz used in the theoretical calculations.}
	
	\begin{figure}[h]
		\centering
		\includegraphics[width=0.7\textwidth]{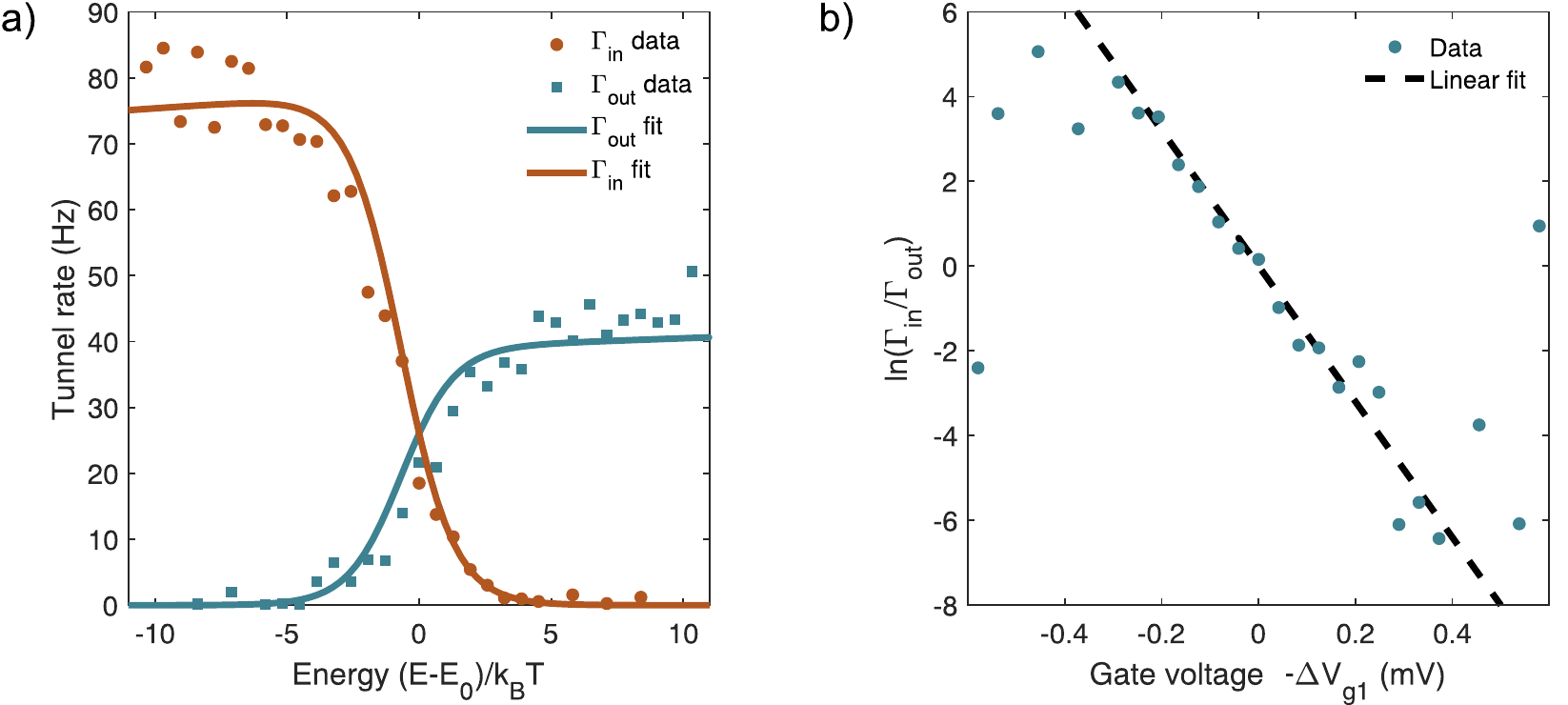}
		\caption{ \marti{a) Solid symbols show the tunnel rates measured at the transition used for the experiment. 
				The detector bias voltage was $V_b = 0.5$ mV and cryostat temperature $T_\mathrm{cryo}=100$ mK. The energy $E_0$ is the energy at which $\Gamma_{\mathrm{in}} = \Gamma_{\mathrm{out}}$, in this case $E_0 = k_BT\ln2$. Solid lines present the fits to the functions $\Gamma_{\mathrm{in}}=\Gamma_0(1+bE)f(E)$ and $\Gamma_{\mathrm{out}}=2\Gamma_0(1+bE)(1-f(E))$ with $\Gamma_0 = 39$~Hz and $b = 0.0036/k_BT$. Here, $f(E)$ is the Fermi-Dirac distribution. b) The dots show the detailed balance plot of the measured rates of panel (a). The dashed line is a linear fit with the slope $-\alpha = -1.6\times10^4\; k_BT/\mathrm{V}$.} }
		\label{fig:tunnelrates}
	\end{figure}
	
	\subsection{RC-filtering in the cryostat lines}
	\noindent \marti{When applying voltage drives with the plunger gates, the signal must pass the cryostat lines which contain an RC filter. The RC filtering distorts the drive shape, but we will show here that this distortion is negligible. In the cryostat used for the experiment, we have $R = 10$~k$\Omega$ and $C = 10$~nF. This gives RC time constant $\tau_\mathrm{RC} = 0.1$~ms, which is two orders of magnitude smaller than the tunneling time scale in our system as shown in the previous section. }
	
\begin{figure}
		\centering
		\includegraphics[width=\textwidth]{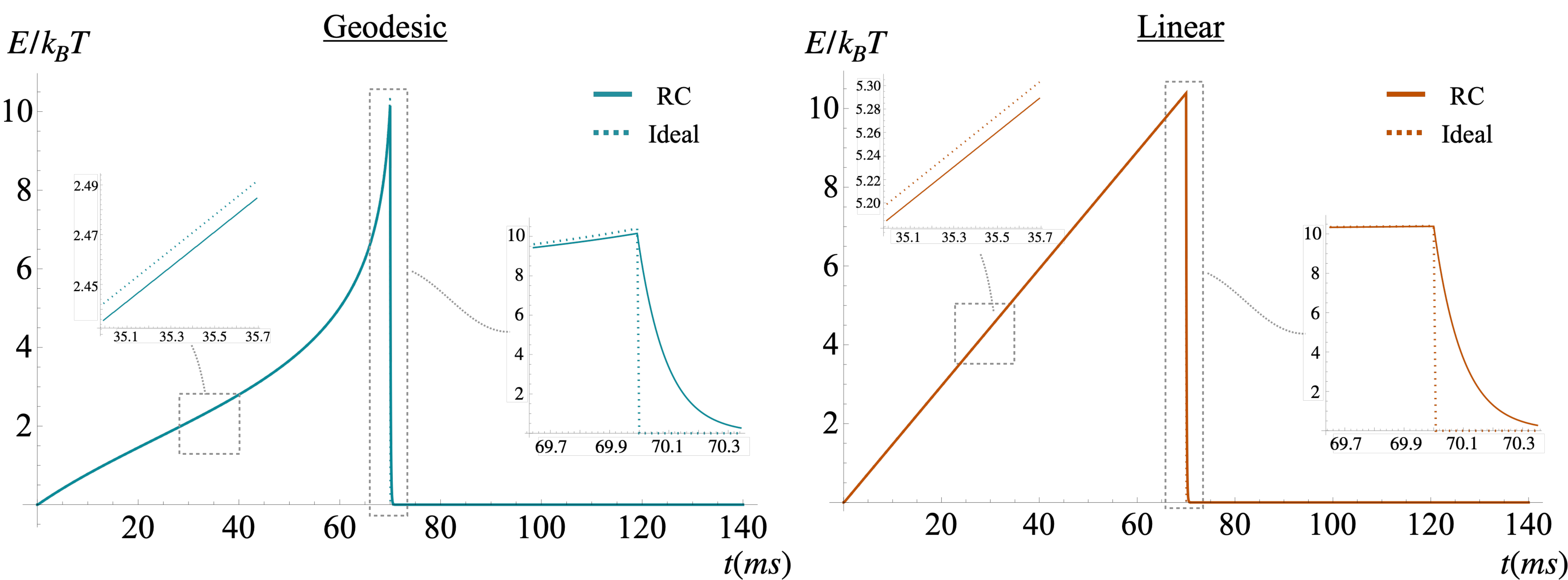}
		\caption{\marti{The effect of the RC filter on the drive shape for the geodesic (left) and the linear drive (right). On the large scale figure of the main panels, the differences to the ideal drive are not visible. The insets show zoomed regions of the plots. The two main differences between the RC-filtered trajectory and the ideal one are: 1) The filtered one lags behind the ideal one by about $\tau_\mathrm{RC} = 0.1$~ms during the ramp, and 2) the final quench has an exponential settling with the timescale $\tau_\mathrm{RC} = 0.1$~ms for the RC filtered drive while the ideal one is infinitely fast.}
		}
		\label{fig:RCdrive}
	\end{figure}

	\marti{The largest distortions to the drive arise with the fast drive with duration $\tau = 70$~ms. Supplemental Figure~\ref{fig:RCdrive} presents the ideal drive and the distorted one after applying the RC filter for this case. Using the RC filtered drive in calculating the the dissipated heat yields a difference of less than $0.01k_BT$ compared to the ideal drive results in the main text. These differences are much smaller than the deviance between experiment and theory in the main letter and hence the finite risetime effects are neglected in the analysis.}
	
	\subsection{Temperature of the electronic reservoir}
	\noindent \marti{To perform the optimal erasure protocol, the system tunneling rates need to be set by the electronic bath and its temperature $T$. To ensure this and that the electronic bath is well thermalized, we use the approach of Refs.~\citealp{Saira2010,Saira2012,Kung2012} and perform the measurements at an elevated cyostat temperature $T_\mathrm{cryo}$ so that the electronic bath temperature follows it, i.e. $T = T_\mathrm{cryo}$.
		Towards this end, we first determine the lever arm $\alpha$ in units of eV/V instead of the $k_BT/$V units relevant for the main study. The energy reference for this measurement is obtained by applying a $1.5$ mV bias voltage over the double quantum dot. The bias voltage opens up triangular 1.5 meV energy windows to the measured hexagonal charge stability diagram of Supplemental Fig.~\ref{fig:backaction2} (a). 
		With the approach of Ref.~\citealp{Taubert2011}, the size of these triangles yields $\alpha = 0.18$ eV/V, which we then use to determine the temperature $T = 0.18/1.6\times10^4\; eV/k_B = 130$ mK of the detailed balance measurement of Supplemental Fig.~\ref{fig:tunnelrates} (b). Supplemental Figure~\ref{fig:backaction2}~(b) repeats the temperature determination procedure for varying cryostat bath temperature $T_{\mathrm{cryo}}$, with Supplemental Fig.~\ref{fig:backaction2} (c)  showing the detailed balance measurement at a few of the bath temperatures $T_\mathrm{cryo}$. }
	
	\marti{For $T_\mathrm{cryo} > 50$ mK, the two temperatures coincide within experimental accuracy of $\sim 10 - 20 \%$ in the determination of the finite bias triangles. This implies that the relevant temperature in the tunnel rates $\Gamma_{in}$ and $\Gamma_{out}$ is set by the cryostat bath temperature, meaning that the electronic reservoir is properly thermalized to the bath and this sets the energy distribution in the tunneling rates. The saturation to $T\approx 50$ mK for $T_\mathrm{cryo} < 50$ mK is the typical saturated electronic temperature for the cryostat used in the experiments. The erasure experiment in the main text was performed at $100$ mK in order to be well above the saturation regime. Since this calibration measurement was done after a thermal cycle of the cryostat, the electron occupancy of the quantum dot is not exactly the same as in the main text. Therefore, the prefactor $\Gamma_0$ in the tunneling rates was roughly twice as large compared to the case in Supplemental Figure~\ref{fig:tunnelrates} (a).}

	\begin{figure}[t]
		\centering
		\includegraphics[width=\textwidth]{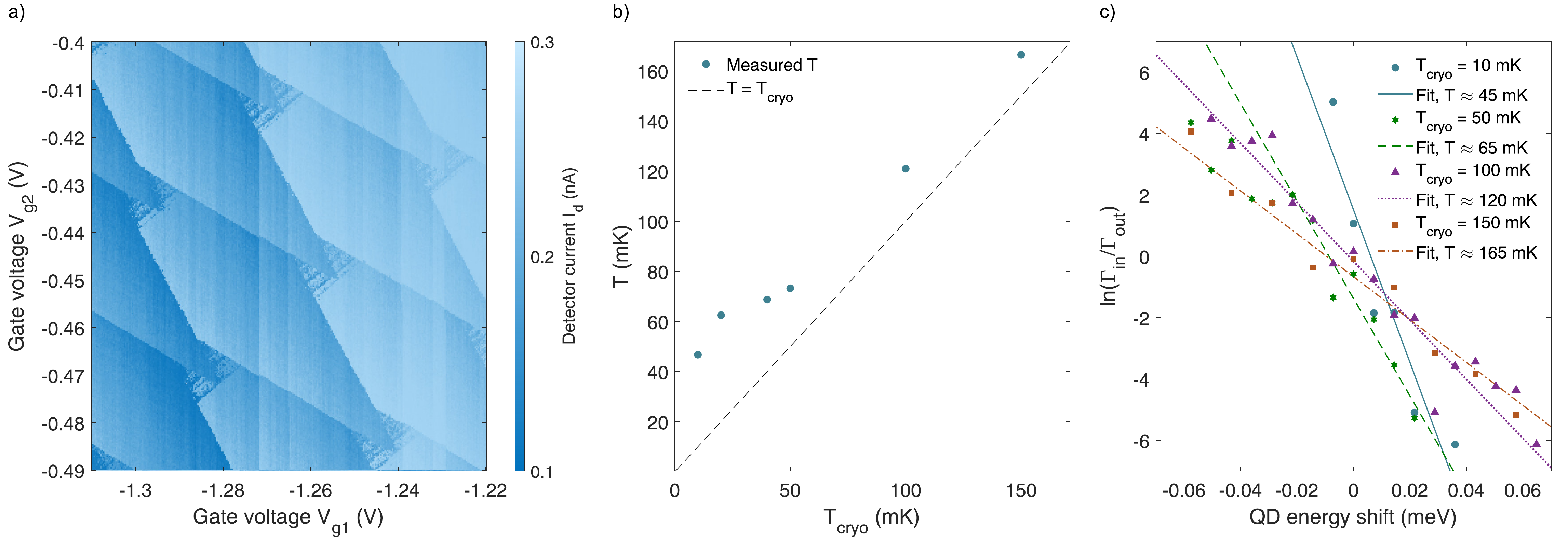}
		\caption{ \marti{ a) Measured detector current $I_\mathrm{d}$ as a function of the plunger gate voltages $V_{g1}$ and $V_{g2}$ of the double quantum dot at finite bias of 1.5 mV applied across the double dot. b) Experimentally determined electronic reservoir temperature $T$ as a function of the cryostat thermometer temperature $T_\mathrm{cryo}$. The dashed line indicates equality $T = T_\mathrm{cryo}$. The difference from the diagonal at the higher ($\geq 100$~mK) $T_\mathrm{cryo}$ is within the precision of determining $\alpha$ from the finite bias triangles. c) The detailed balance measurement similar to the one in Supplemental Fig.~\ref{fig:tunnelrates} (b) for $T_\mathrm{cryo}=10$ mK, $50$ mK, $100$ mK, and $150$ mK. The lines show fits to the linear part which was used to determine the temperature $T$ in panel (b).} }
		\label{fig:backaction2}
	\end{figure}
	
	\subsection{Minimizing the detector backaction}
	\noindent \marti{In addition to performing the experiment at elevated temperature, we found that it is also necessary to select carefully the operation point of the detector in order to minimize detector back-action. Supplemental Figure~\ref{fig:lowbias} (a) presents the same measurement as in Supplemental Fig.~\ref{fig:tunnelrates} (a) but at lower detector bias voltage of $V_b =~0.1$ mV instead of the $V_b =~0.5$ mV used in the actual study. Supplemental Figure~\ref{fig:lowbias} (b) shows a $5$~s time trace of the detector current for both of these setpoints where $\Gamma_\mathrm{in}\approx\Gamma_\mathrm{out}$. In the ideal case, the tunnel rates $\Gamma_\mathrm{in}$ and $\Gamma_\mathrm{out}$ both should reach half of their corresponding maximum values of $\Gamma_0$ and $2\Gamma_0$ at the energy $E = 0$ corresponding to the Fermi level of the reservoir. However, in Supplemental Fig.~\ref{fig:lowbias} (a) we see that these half way points marked with the dashed lines are separated by $\sim 1.8k_BT$. This is an increase of $\sim 1.5k_BT$ compared to Supplemental Figure~\ref{fig:tunnelrates}~(a).}
	
	\begin{figure}
		\centering
		\includegraphics[width=\textwidth]{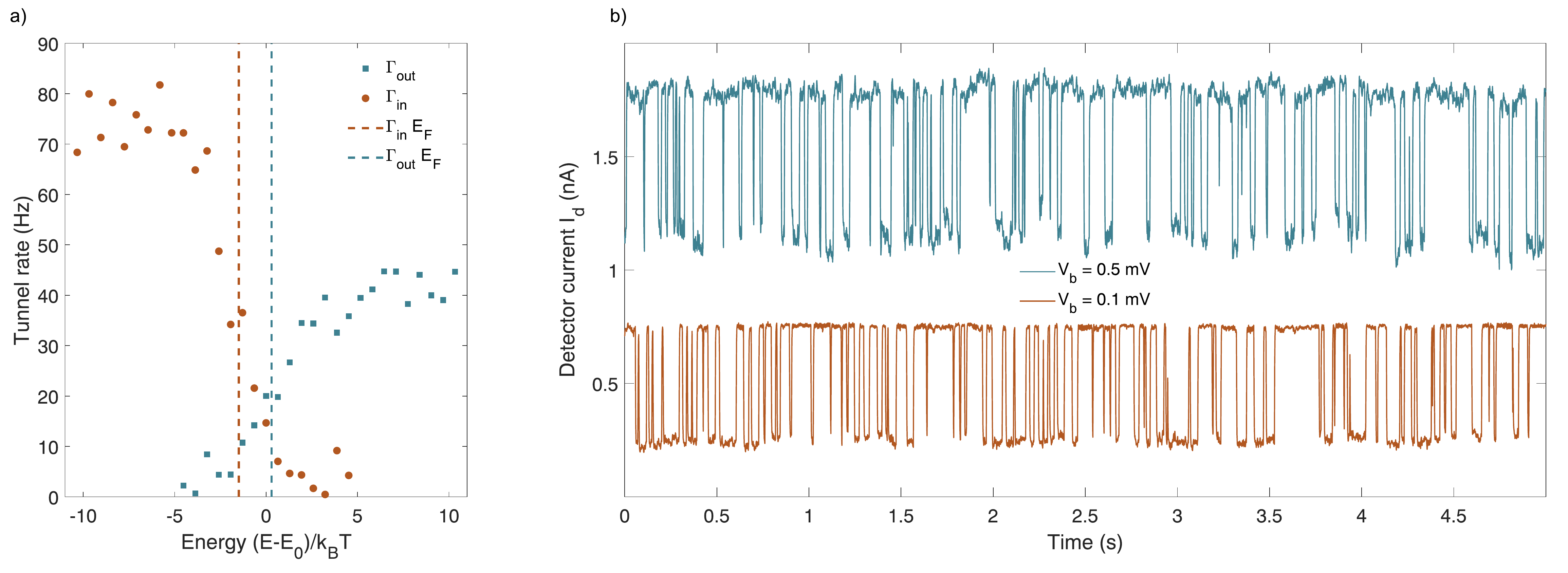}
		\caption{ \marti{a) The tunnel rate measurement similar to Supplemental Figure~\ref{fig:tunnelrates}, but measured with a detector bias voltage of $0.1$ mV. The dashed lines indicate the voltages where the rates reach half of their saturated value. These points are here separated by $\sim 1.8k_BT$, suggesting the QD energy level is shifted when a tunneling event happens. b) Time traces of the detector current at $\Gamma_\mathrm{in} \approx \Gamma_\mathrm{out}$ for the detector bias voltages of $V_b = 0.5$~mV (top trace) and $0.1$~mV (bottom trace). In the top trace, the $n=1$ state has a detector current about $1.5$ times larger than the $n=0$ state whereas in the bottom trace, it is $4$ times larger.}}
		\label{fig:lowbias}
	\end{figure}
	
	\marti{Our understanding of this Fermi energy shift is that the detector acts back to the system and gates it by a small amount: In the same way as the occupancy of the system moves the detector QD energy level, the occupancy $n_\mathrm{det}$ of the detector shifts the energy position of the system. Comparing the two setpoints, we see that the lower bias results in a relative signal strength of $\sim 0.8~\mathrm{nA} / 0.2~\mathrm{nA} = 4$, while the corresponding metric for the higher bias setpoint is only $1.5$. Our interpretation is that this increase in relative signal strength comes with a corresponding increase in back-action on the system. 
		In the experiment in the main text, we used the higher $V_b$ to minimize the energy shift.}

	\section{Irreversibility and thermodynamic length}\label{app:theory}
	Heat can be split in a path independent contribution, proportional to the difference in entropy at the endpoints, and a path dependent term, that goes under the name of dissipation. In this appendix we show how one can derive this splitting and we sketch how this can be used to derive a metric structure on the space of parameters. Finally, we analyse the example of a two level system. The derivations mirror the one provided in~\cite{scandiThermodynamicLengthOpen2019,abiusoGeometricOptimisationQuantum2020}, to which we refer for further details.
	
	\subsection{General theory}
	
	Consider a generic quantum state $\rho (t)$ undergoing an open system dynamics, to which we can associate a driven system Hamiltonian $H(t)$. The corresponding thermal state at each moment is denoted by $\rho_{eq} (t) := \frac{e^{- H(t)/{k_BT}}}{\mathcal{Z} (t)}$,  where $\mathcal{Z} (t):= \Tr{e^{- H(t)/k_BT}}$ is called partition function. We assume that for each $t$ the dynamics tries to equilibrate the state to $\rho_{eq} (t)$. Now, simply by using the functional form of the thermal distribution, we can rewrite the average heat as:
	\begin{align}
		\average{\Delta Q}{} &= -\int_0^\tau \dt\, \Tr{\dot \rho(t) H(t)} = k_BT\int_0^\tau \dt\, \Tr{\dot \rho(t) \log e^{- H(t)/k_BT}} =\\
		&=k_BT\int_0^\tau \dt\, \Tr{\dot \rho(t) \log \rho_{eq}(t)},\label{eq:A2}
	\end{align}
	where in the second line we exploited the fact that the trace condition for $\rho (t)$ implies that $\Tr{\dot\rho(t)}=0$, so we can complete the thermal state. 
	
	With the hindsight of Clausius' inequality, we add and subtract the total derivative of the entropy to Eq.~\eqref{eq:A2}, which leads to:
	\begin{align}
		\average{\Delta Q}{} &= k_BT\int_0^\tau \dt\, \norbra{\frac{\de}{\dt}\Tr{ \rho(t) \log \rho (t)} - \frac{\de}{\dt}\Tr{ \rho(t) \log \rho (t)}+ \Tr{\dot \rho(t) \log \rho_{eq}(t)}} = \\
		&= -k_BT \,\Delta S + k_BT \int_0^\tau \dt\, (-\partial_{\rho(t)} S(\rho(t)||\rho_{eq} (t))),\label{eq:A4}
	\end{align}
	where $S(\rho||\sigma) : = \Tr{\rho(\log\rho - \log\sigma)}$ is the usual relative entropy, and we used the notation:
	\begin{align}
		\partial_{\rho(t)} S(\rho(t)||\rho_{eq} (t)) := \lim_{\varepsilon\rightarrow 0} \frac{S(\rho(t+\varepsilon)||\rho_{eq} (t))-S(\rho(t)||\rho_{eq} (t))}{\varepsilon},
	\end{align}
	for the partial derivative of $S(\rho(t)||\rho_{eq} (t))$ with respect to the first argument only (i.e., keeping the instantaneous thermal state fixed during the differentiation). In this way, not only we provide a derivation of the Clausius' statement of the second law ($\average{\Delta Q}{}+ k_BT\Delta S\geq 0$), but we also obtain an explicit expression for the dissipation. In particular, it can be shown that the integral in Eq.~\eqref{eq:A4} is always positive for evolutions induced by quantum channels~\cite{breuerTheoryOpenQuantum2009,scandiThermodynamicLengthOpen2019}. Moreover, if the driving is infinitesimally slow, the system is always at equilibrium; then, it is apparent that the dissipation is zero, giving the equality $\average{\Delta Q}{} =- k_BT\Delta S$. 
	
	In fact, for Markovian evolutions, this is the only case in which the dissipation is zero. The idea behind thermodynamic length is to expand the dissipation around this global minimum. That is, we consider protocols that are realised in a time $\tau$ much bigger than any equilibration timescale of the system (but still finite). For definiteness, and without loss of generality, we rewrite the Hamiltonian as $H(t) = \sum_i \lambda^i (t)  X_i$, where  $\{\lambda^i (t) \}$ are scalars representing the time-dependent externally controllable parameters, and $\{X_i\}$ are the corresponding observables. Then, it can be shown that the dissipation in the slow driving regime takes the form:
	\begin{align}
		\average{\Delta Q}{} + k_BT\Delta S= k_BT \int_0^\tau \dt\,\dot\lambda^i(t)\dot\lambda^j(t)\, g_{i,j}(t) +\bigo{\frac{1}{\tau^2}}\label{eq:A6}
	\end{align}
	where $g_{i,j}(t)$ is a positive, symmetric form, which depends smoothly on the base-point~\cite{scandiThermodynamicLengthOpen2019,abiusoGeometricOptimisationQuantum2020}. Intuitively, $g_{i,j}(t)$ can be interpreted as the Hessian of the dissipation around the minimum given by setting $1/\tau = 0$, (that is for infinite duration of the protocol). This explains the positivity and the symmetry.
	
	Thanks to these conditions, $g_{i,j}(t)$ naturally induces a metric structure on the space of parameters. The integral in Eq.~\eqref{eq:A6} is a standard object in this context, called energy functional, and it can be minimised by solving the corresponding geodesics equation. This provides an automatic method for finding optimal protocols (i.e., minimally dissipating) without having to resort to anything more complicated than the solution of a system of second order differential equations. 
	
	\subsection{Quantum dot in contact with a bath: simplified version}
	
	We exemplify the construction above for a quantum dot whose excited state evolves according to the rate equation:
	\begin{align}
		\average{\dot n(t)}{} = \Gamma_0 (n_{eq}(E(t))- \average{ n(t)}{}),
	\end{align}
	where $E(t)$ is a fixed (but arbitrary) driving. This dynamics is a slight variation of the rate equation describing the experiment, but we chose this expression to obtain analytical results. It can be proven by perturbative methods, or through numerical simulations, that the results are not very sensitive to this change (as it can be noticed by comparing the figures presented here to the one in the main text). We further comment on the differences between this simplified version and the realistic model in the next subsection for what regards the metric and the Christoffel symbol.
	
	Let us see how the dynamics of $\average{ n(t)}{}$ is affected by a change in the duration $\tau$ of the protocol. First, it is apparent that in the limit $\tau\rightarrow\infty$ the driving appears frozen to the system, so the population is exactly given by the one of the excited state $\average{ n(t)}{} = n_{eq}(E(t))$.  
	
	\begin{figure}
		\centering
		\includegraphics[width=0.7\columnwidth]{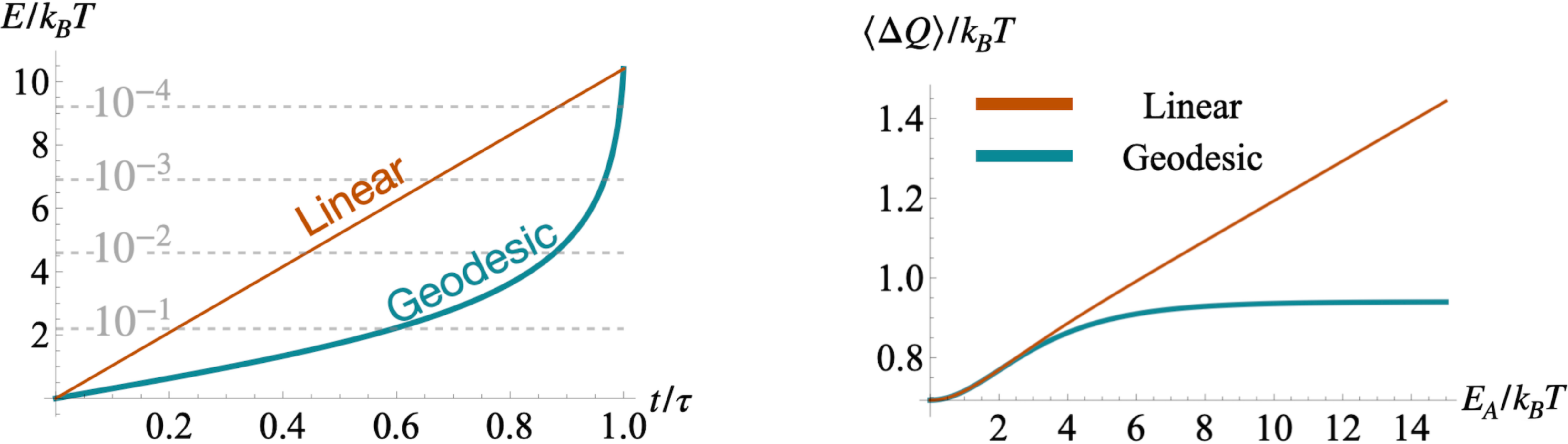}
		\caption{On the left, representation of the geodesic for a qubit, whose analytical expression is given in Eq.~\eqref{eq:A16}. On the right, average heat produced in a protocol to cancel a qubit as a function of the driving amplitude. Here we only represent the first order contribution to the dissipation (i.e., Eq.~\eqref{eq:A19} and Eq.~\eqref{eq:A21}) so the effect appears to be more stark than the one presented in the main text. Still, one can obtain the same behaviour by extending the duration of the experiment.}
		\label{fig:figApp}
	\end{figure}
	
	Define the thermalisation timescale $\tau_{eq} := \Gamma_0^{-1}$. When we talk about slow driving we always mean that $\tau_{eq}/\tau \ll 1$. In this regime, we can perturbatively solve the expression of $\average{ n(t)}{}$. We rewrite the population as:
	\begin{align}
		\average{ n(t)}{} = n_{eq}(E(t)) + \Delta \,n(t).
	\end{align} 
	where $\Delta \,n(t)$ is of order $\bigo{\tau_{eq}/\tau}$. Plugging this ansatz into the rate equation we obtain:
	\begin{align}
		&\frac{\de}{\dt}n_{eq}(E(t))  = - \Gamma_0 \Delta \,n(t) + \bigo{\norbra{\frac{\tau_{eq}}{\tau}}^2}\implies\\
		\implies&\Delta \,n(t) = \frac{\tau_{eq}}{2}\frac{\dot E(t)}{1+ \cosh (E(t)-\log 2)}
	\end{align}
	where we kept only the thermal state in the right hand side, as every differentiation increases one order in $\tau_{eq}/\tau$. Then, the dissipation takes the form:
	\begin{align}
		\average{\Delta Q}{} + k_BT\Delta S&= k_BT \int_0^\tau \dt\, (-\partial_{\rho(t)} S(\rho(t)||n_{eq} (E(t)))) =\\
		&= k_BT\,\tau_{eq} \,\int_0^\tau \dt\,  \frac{\dot E(t)^2}{2+ 2\cosh (E(t)-\log 2)} + \bigo{\norbra{\frac{\tau_{eq}}{\tau}}^2},\label{eq:A12}
	\end{align}
	where we omitted the lengthy but straightforward calculations. In this case the metric is given by:
	\begin{align}
		g (t) = \frac{1}{2+ 2\cosh (E(t)-\log 2)},
	\end{align}
	where we dropped the indices, since there is only one parameter. The corresponding Christoffel symbol can be computed as:
	\begin{align}
		\Gamma(t) = \frac{1}{g (t)}\, \frac{\de g (t)}{\de E(t)}  = -\frac{1}{2} \tanh \left(\frac{E(t)-\log 2}{2}\right),
	\end{align}
	and the geodesic equation is given by:
	\begin{align}
		\ddot E(t) + \Gamma(t) \dot E(t)^2 = 0.
	\end{align}
	This can be analytically solved to give a closed form for the optimal protocol:
	\begin{align}
		&E_{g}(t) = 2 \log \left(\sqrt{2}\cot \left(A + B \frac{t}{\tau}\right)\right),\label{eq:A16}
	\end{align}
	where $A$ and $B$ can be chosen to fix the initial and final energy $E_0$,  $E_1$. In particular, choosing $E_0 = \log 2$, and $E_1 = E_0 + E_A$, the two constants take the value $A = \pi/4$ and $B = -\frac{1}{2} \arcsin( \tanh \frac{E_A}{2})$. 
	
	There is an interesting property that geodesics satisfy in general: they keep the entropy production rate constant. This means that the integrand in Eq.~\eqref{eq:A12} does not depend on time, as it can be checked by direct calculation:
	\begin{align}
		\frac{\dot E_g(t)^2}{2+ 2\cosh (E_g(t)-\log 2)} = \frac{4\,B^{2}}{\tau^{2}} = \frac{1}{\,\tau^{2}}\norbra{\arcsin \tanh \frac{E_A}{2}}^{2},
	\end{align}
	hence the total dissipation can be easily computed to be:
	\begin{align}
		\average{\Delta Q}{}+ k_BT\Delta S&= k_BT\,\frac{\tau_{eq}}{\,\tau^{2}} \int_0^\tau \dt\norbra{\arcsin \tanh \frac{E_A}{2}}^{2} +\bigo{\norbra{\frac{\tau_{eq}}{\tau}}^2}=\\
		&= k_BT\,\frac{\tau_{eq}}{\,\tau}\norbra{\arcsin \tanh \frac{E_A}{2}}^{2} +\bigo{\norbra{\frac{\tau_{eq}}{\tau}}^2}.\label{eq:A19}
	\end{align}
	This should be compared with the behaviour for a linear drive, for which one obtains:
	\begin{align}
		\average{\Delta Q}{} + k_BT\Delta S&= k_BT\,\frac{\tau_{eq}}{\tau} \,\int_0^\tau \frac{\dt}{\tau}\,  \frac{ E^2_A}{2+ 2\cosh (E_A\,\frac{t}{\tau})} + \bigo{\norbra{\frac{\tau_{eq}}{\tau}}^2}=\\
		&= k_BT\,\frac{\tau_{eq}}{\,\tau} \norbra{\frac{E_A}{2}\tanh \frac{E}{2}}+ \bigo{\norbra{\frac{\tau_{eq}}{\tau}}^2}.\label{eq:A21}
	\end{align}
	As it is shown in Fig.~\ref{fig:figApp}, where we plotted the total heat in the two cases as a function of the amplitude, for the geodesic drive the dissipation saturates to $\pi^2/4$, while for the linear drive it linearly diverges. This behaviour mirrors the one shown in Fig.~4 in the main text.
	
	\subsection{Quantum dot in contact with a bath: realistic model}
	
	In the experiment the rate equation is given by:
	\begin{align}
		\average{\dot n(t)}{} = \Gamma_0 (1+b E(t))(1+f(E(t))) (n_{eq}(E(t))- \average{ n(t)}{}).\label{app:eq:A22}
	\end{align}
	Reproducing the considerations of the previous section, we can postulate the ansatz $\average{ n(t)}{} = n_{eq}(E(t)) + \Delta \,n(t)$, where in this case the perturbation is given by:
	\begin{align}
		\Delta \,n(t) = \frac{\tau_{eq}}{2}\frac{\dot E(t)}{(1+b E(t))(1+f(E(t))) (1+ \cosh (E(t)-\log 2))}.
	\end{align} 
	Then, Eq.~\eqref{eq:A12} becomes in this context 
	\begin{align}
		\average{\Delta Q}{} + k_BT\Delta S&= k_BT\,\tau_{eq} \,\int_0^\tau \dt\,  \frac{\dot E(t)^2}{(1+b E(t))(1+f(E(t))) (2+ 2\cosh (E(t)-\log 2))} + \bigo{\norbra{\frac{\tau_{eq}}{\tau}}^2},
	\end{align}
	so that the metric is given by:
	\begin{align}
		g (t) = \frac{1}{(1+b E(t))(1+f(E(t))) (2+ 2\cosh (E(t)-\log 2))},
	\end{align}
	while the Christoffel symbol reads:
	\begin{align}
		\Gamma(t) = \frac{1}{2} \left(-\frac{b}{b E(t)+1}+\frac{1}{e^{-E(t)}+1}+\frac{6}{e^{E(t)}+2}-2\right).
	\end{align}
	Despite the lengthy expressions of these quantities, the effective difference between this model and the one in the previous section are quite small. 
	
	\section{Numerical simulations}	
	\label{App:NumericalSimulations}
	
	
	\subsection{From slow to fast driving}
	
\begin{figure}
		\centering
		\includegraphics[width=.7\columnwidth]{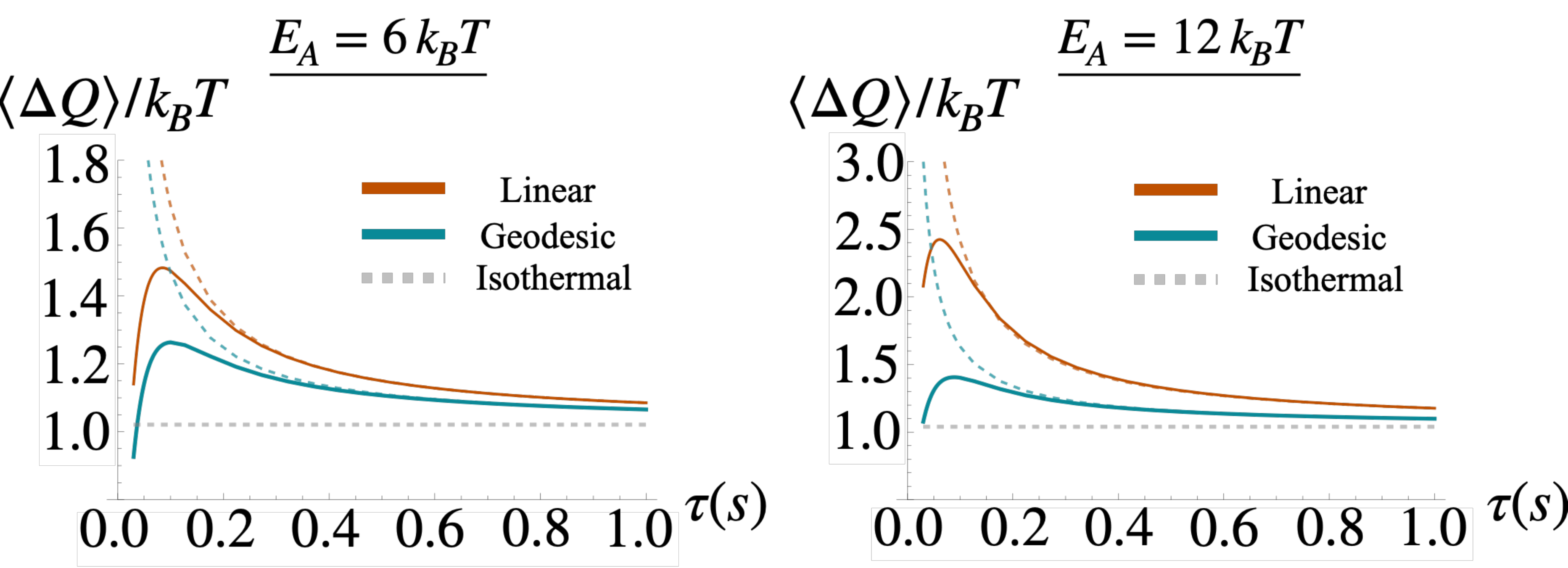}
		\caption{\marti{Numerical simulations of the heat produced for the two protocols considered in the main text as a function of $\tau$ for $E_A=6\,k_B T$ and $E_A=12\,k_B T$. The solid lines correspond to actual simulations, while the dashed lines come from the slow driving approximation.}  }
		\label{fig:newFigApp}
	\end{figure}

	\marti{
		In the main text, we show experimental results for $\tau=1.0\,s$, (with $\Gamma_0 \tau\simeq 40$ hence corresponding to the slow driving regime), and for $\tau=0.07\,s$  ($\Gamma_0 \tau=3$ and hence relatively fast driving compared to the relaxation timescale). Here we complement these considerations with numerical simulations for a whole range of $\tau\in [0,1]\,s$, as shown in Fig.~\ref{fig:newFigApp}. In there, we compare the linear protocol with the geodesic one, together with the slow driving approximation (coming from a perturbative expansion in which terms of order $\mathcal{O}((\Gamma_0 \tau)^{-2})$ and higher are neglected). From this plot we can see that   $\tau=0.07\,s$ is clearly away from the slow driving approximation, whereas $\tau=1.0\,s$ is deep in the slow driving regime. Remarkably, the geodesic protocol, which is derived in the slow driving regime, shows noticeable advantages compared to the linear one in the whole range of $\tau$s, further supporting our claims in the main text.  
	}

	\subsection{Geodesic erasure with a transient}

	In the main text it is argued that allowing the system to equilibrate at the end of the protocol would make the geodesic drive optimal both in terms of minimal dissipation and erasure quality even for fast drivings. We present here numerical evidences that this is in fact the case.
	
	\begin{figure}
		\centering
		\includegraphics[width=0.7\columnwidth]{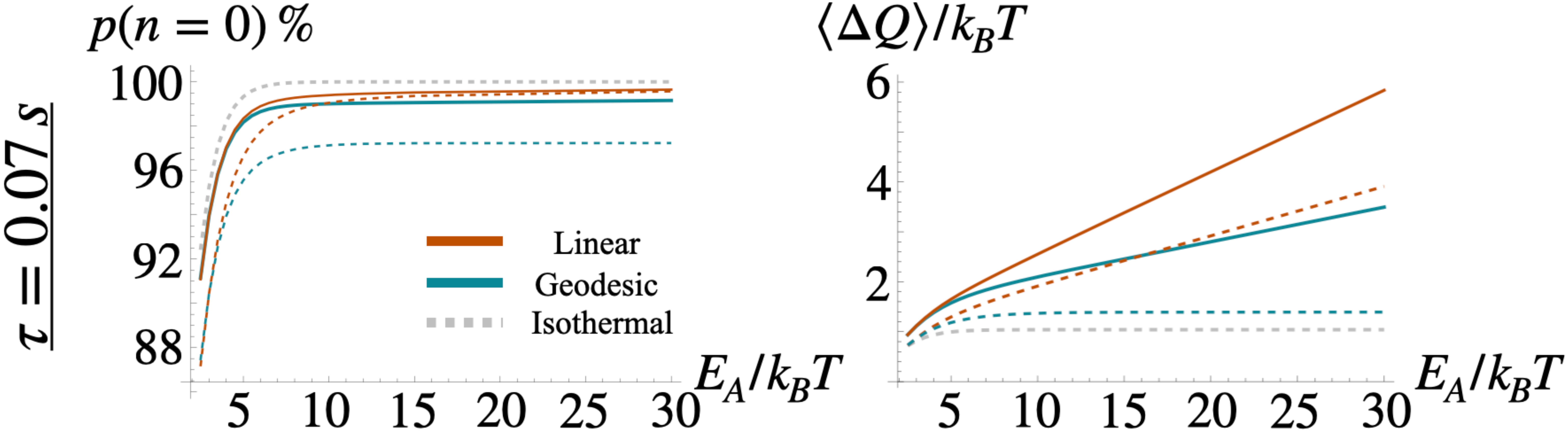}
		\caption{Numerical simulations showing how allowing time for an extra transient can improve the quality of erasure of the geodesic protocol, while still making it preferable to the linear drive in respect to the amount of dissipated heat. The solid lines correspond to the drivings in which the system is driven for time $\tau$ and then it is let thermalise for an extra time $\tau_{\rm trans}$, while the dashed lines correspond to allocating time $\tau+\tau_{\rm trans}$ to the drive.}
		\label{fig:figApp2}
	\end{figure}
	
	To this end, we consider a transient time $\tau_{\rm trans}$ of twice $\Gamma_0(1+b E_1)(1+f(E_1))$ (the prefactor in Eq.~\eqref{app:eq:A22}), which in our experimental set-up corresponds to~$\sim 0.05 \,s$. There are two ways of allocating the total time $\tau+\tau_{\rm trans}$: either the full time is used for the drive, or the drive takes place in time $\tau$, while for the remaining $\tau_{\rm trans}$ the system thermalises at fixed energy. These two possible choices are shown in Fig.~\ref{fig:figApp2}: the dashed lines correspond to the first choice, while the solid lines correspond to the latter.
	
	The solid lines show how the extra thermalisation makes the erasure quality of the geodesic and linear drive comparable (the difference is of order $\sim0.1\%$) while the dissipation for the geodesic is always lower in the first case. On the other hand, when considering also the red dashed line (the one corresponding to a linear drive which takes $\tau+\tau_{\rm trans}$) there is a trade-off: for lower amplitudes it dissipates less than  the geodesic, but at the cost of a lower erasure quality; for higher amplitudes, instead, the geodesic starts dissipating less, whereas the erasure quality saturates to a similar value. This shows that even if the geodesic drive is in principle designed only for slowly driven systems, it can be minimally modified to give drives that are optimal also in the fast driving regime.

\end{document}